\documentclass[11pt]{article}
\usepackage[dvips]{graphicx}
\usepackage{amssymb}

\title{Structural distortions 
       of frustrated quantum spin lattices
       \protect\\
       in high magnetic fields}
\author{Oleg Derzhko$^{1,2}$,
        Johannes Richter$^2$
        and
        J\"{o}rg Schulenburg$^3$\\
\small{$^1$Institute for Condensed Matter Physics,
       National Academy of Sciences of Ukraine,}\\
\small{1 Svientsitskii Street, L'viv-11, 79011, Ukraine}\\
\small{$^2$Institut f\"{u}r Theoretische Physik,
       Universit\"{a}t Magdeburg,}\\
\small{P.O. Box 4120, D-39016 Magdeburg, Germany}\\        
\small{$^3$Universit\"{a}tsrechenzentrum,
       Universit\"{a}t Magdeburg,}\\
\small{P.O. Box 4120, D-39016 Magdeburg, Germany}}

\date{\today}

\begin{document}

\renewcommand\baselinestretch {1.4}
\large\normalsize

\maketitle

\begin{abstract}
We study the stability 
of some strongly frustrated antiferromagnetic spin lattices 
in high magnetic fields 
against lattice distortions.
In particular, 
we consider a spin-$s$ anisotropic Heisenberg antiferromagnet 
on the square-kagom\'{e} and kagom\'{e} lattices.
The independent localized magnons 
embedded in a ferromagnetic environment, 
which are the ground state at the saturation field,
imply lattice instabilities  
for appropriate lattice distortions
fitting to the structure of the localized magnons.
We discuss in detail 
the scenario of this spin-Peierls instability in high magnetic fields 
which essentially depends on the values 
of the exchange interaction anisotropy $\Delta$ and spin $s$.
\end{abstract}

\vspace{2mm}

\noindent
{\bf PACS number(s):}
75.10.Jm, 75.45.+j

\vspace{2mm}


\vspace{5mm}

\renewcommand\baselinestretch {1.55}
\large\normalsize

\section{Introduction. Localized magnons}
\label{s1}

The square-kagom\'{e} and kagom\'{e} Heisenberg antiferromagnets 
have attracted much interest in recent times 
because of interplay between quantum fluctuations and frustrated lattice structure 
\cite{a,b,01,02,03}.
Recently,
for a wide class of frustrated spin lattices,
which includes both the square-kagom\'{e} and kagom\'{e} lattices,
exact eigenstates consisting of independent localized magnons
in a ferromagnetic environment have been found \cite{d,04,05}.
They become ground states if the saturating magnetic field is applied
and lead to a macroscopic jump
in the zero-temperature magnetization curve just below saturation.
Moreover,
very recently
we have examined 
a field-tuned instability 
of the square-kagom\'{e} and kagom\'{e} spin lattices
with respect to  lattice distortions 
through a magnetoelastic mechanism \cite{c}
reporting rigorous  analytical results
completed by large-scale exact diagonalization data
for lattices up to $N=54$ sites \cite{06}.
That study was restricted mainly to 
the spin-$\frac{1}{2}$
isotropic Heisenberg antiferromagnet.
Since often one meets kagom\'{e} materials with $s>\frac{1}{2}$
(for instance,
spin-$\frac{3}{2}$ kagom\'{e}-like compound 
Ba$_2$Sn$_2$ZnCr$_{7p}$Ga$_{10-7p}$O$_{22}$
with a comparably small exchange constant of about $37\ldots 40$ K
\cite{e})
it is useful to go beyond the case $s=\frac{1}{2}$ and $\Delta=1$.
In this paper we extend our previous results 
for the spin quantum numbers $s>\frac{1}{2}$ 
and exchange interaction anisotropy  $\Delta\ne 1$
demonstrating that a scenario 
of the discussed earlier magnetic-field induced lattice instability
may essentially depend on the values of $s$ and $\Delta$.

\begin{figure}

\vspace{3mm}

\begin{center}
\includegraphics[clip=on,width=6.0cm,angle=0]{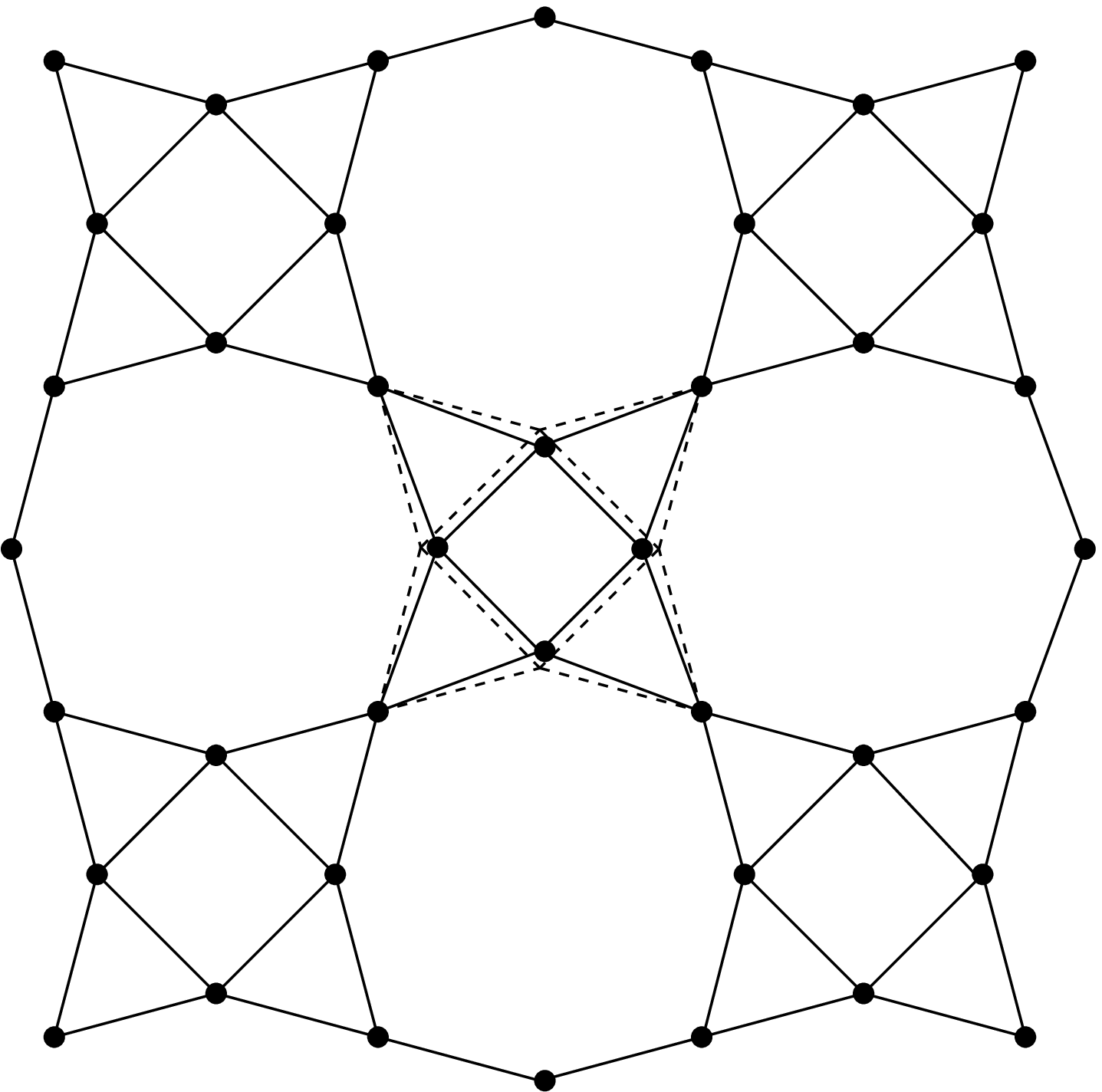}

\vspace{5mm}

\includegraphics[clip=on,width=6.0cm,angle=0]{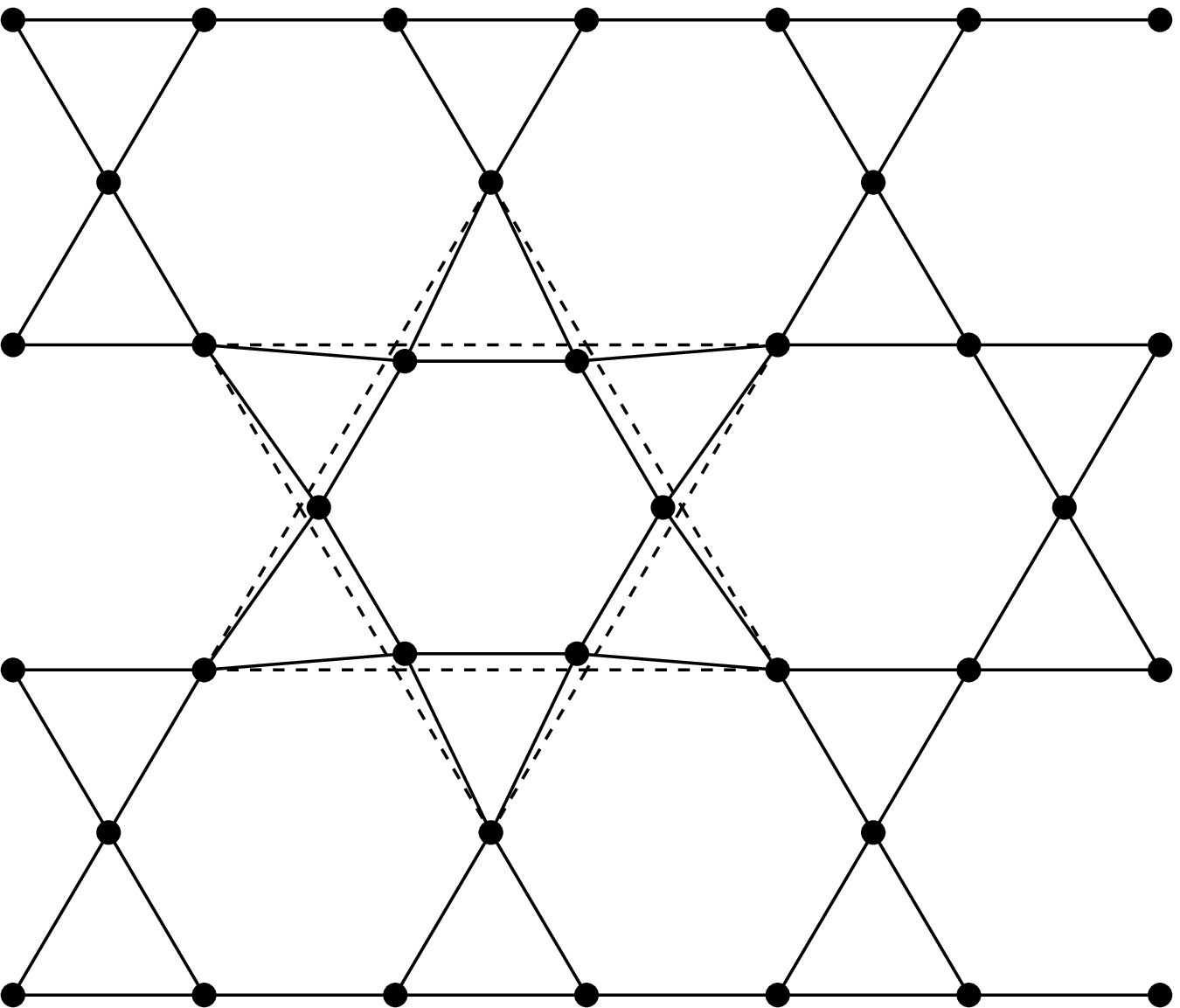}
\end{center}

\vspace{3mm}

\caption[]
{Square-kagom\'{e} lattice with one distorted square
(top)
and
kagom\'{e} lattice with one distorted hexagon
(bottom)
which can host localized magnons.
The parts of the lattices before distortions
are shown by dashed lines.
All bonds in the lattice before distortions have the same length.
\label{fig01}}

\end{figure}

To be specific,
we consider
the square-kagom\'{e} lattice
(Fig. \ref{fig01}, top)
and
the kagom\'{e} lattice
(Fig. \ref{fig01}, bottom).
The Hamiltonian
of $N$ quantum spins
reads
\begin{eqnarray}
\label{01}
H=
\sum_{(nm)}J_{nm}\left(\frac{1}{2}\left(s_n^+s_m^-+s_n^-s_m^+\right)
+\Delta s_n^zs_m^z\right)
-hS^z.
\end{eqnarray}
Here
the sum runs over the bonds (edges)
which connect the sites (vertices)
occupied by  spins
for the lattice under consideration,
$J_{nm}>0$ are the antiferromagnetic exchange constants
between the sites $n$ and $m$,
$\Delta\ge 0$
is the anisotropy parameter,
$h$ is the external magnetic field,
and
$S^z=\sum_ns_n^z$ is the $z$-component of the total spin.
We assume that all bonds in the lattice
without distortion
have the same length
and hence all exchange constants have the same value $J$.

From Refs. \cite{d,04,05,03} we know
that independent localized one-magnon states
embedded in a ferromagnetic background
are exact ground states of the Hamiltonian (\ref{01}) at saturation field 
for the considered models.

More specifically,
by direct computation one can check
that
\begin{eqnarray}
\label{02}
\vert 1\rangle
=
\frac{1}{\sqrt{4}}
\left(\vert s-1,s,s,s \rangle
-\vert s,s-1,s,s \rangle
\right.
\nonumber\\
\left.
+\vert s,s,s-1,s \rangle
-\vert s,s,s,s-1 \rangle\right)
\vert\ldots s \ldots\rangle
\end{eqnarray}
and
\begin{eqnarray}
\label{03}
\vert 1\rangle
=
\frac{1}{\sqrt{6}}
\left(\vert s-1,s,s,s,s,s \rangle
-\vert s,s-1,s,s,s,s \rangle
+\vert s,s,s-1,s,s,s \rangle
\right.
\nonumber\\
\left.
-\vert s,s,s,s-1,s,s \rangle
+\vert s,s,s,s,s-1,s \rangle
-\vert s,s,s,s,s,s-1 \rangle\right)
\vert\ldots s \ldots\rangle
\end{eqnarray}
are one-magnon eigenstates of the Hamiltonian (\ref{01})
on the square-kagom\'{e} and kagom\'{e} lattices.
Here 
$s$ or $s-1$ denote the value of $s_n^z$, 
the omitted site indices in the first multiplier 
in the r.h.s. of Eq. (\ref{02}) 
(Eq. (\ref{03}))
run along a square (hexagon) trapping cell,
and
the second multiplier in the r.h.s. of Eqs. (\ref{02}) and (\ref{03}),
$\vert\ldots s \ldots\rangle$,
stands for the embedding fully polarized ferromagnetic environment.
The corresponding energies 
($h=0$) of the one-magnon states (\ref{02}) and (\ref{03}) are
\begin{eqnarray}
\label{04}
E_1
=
-2sJ +2s(2s-1)\Delta J
+2s(4s-1)\Delta J
+(2N-12)s^2\Delta J
\end{eqnarray}
and
\begin{eqnarray}
\label{05}
E_1
=
-2sJ+2s(3s-1)\Delta J
+2s(6s-1)\Delta J
+(2N-18)s^2\Delta J.
\end{eqnarray}
We separate in Eqs. (\ref{04}), (\ref{05})
the contributions to the energy
from those bonds which form a magnon trapping cell (first and second terms),
from the bonds connecting this cell with the environment (third terms)
and
from the ferromagnetic environment (fourth terms).
The considered lattices may contain $n=1,\ldots,n_{\max}$ localized magnons
where 
$n_{\max}=\frac{1}{6}N$
or
$\frac{1}{9}N$
for the square-kagom\'{e} and kagom\'{e} lattices,
respectively.
Each magnon decreases the total $S^z$ by one
and a localized magnon state with $n$ independent magnons
has $S^z=sN-n$. 
In the presence of an external field $h\ne 0$
the energy $E(S^z,h)$
can be obtained from the energy without field $E(S^z)$
through the relation
$E(S^z,h)=E(S^z)-hS^z.$

Under quite general assumptions it was proved \cite{d,07}
that the localized magnon states have lowest energies
in the corresponding sectors of total $S^z$.
As a result,
these states
become ground states
at the saturation field.
More specifically,
the ground-state energy in the presence of a field
is given by
$E_0(h)=E_{\min}(S^z)-hS^z$
and the ground-state magnetization $S^z$
is determined from the equation
$h=E_{\min}(S^z)-E_{\min}(S^z-1)$.
Since for $S^z=sN,\ldots,sN-n_{\max}$ 
the localized magnon states
are the lowest states, one has
\begin{eqnarray}
\label{06}
E_{\min}(S^z)
=2Ns^2\Delta J-n\epsilon_1
=2Ns^2\Delta J-Ns\epsilon_1+\epsilon_1S^z
\end{eqnarray}
where
\begin{eqnarray}
\label{07}
\epsilon_1
=2s(1-(2s-1)\Delta)J-2s(4s-1)\Delta J+12s^2\Delta J
=2s(1+2\Delta)J
\end{eqnarray}
and
\begin{eqnarray}
\label{08}
\epsilon_1
=2s(1-(3s-1)\Delta)J-2s(6s-1)\Delta J+18s^2\Delta J
=2s(1+2\Delta)J
\end{eqnarray}
for the square-kagom\'{e} and kagom\'{e} lattices,
respectively.
Due to the linear relation between $E_{\min}$ and $S^z$ (\ref{06})
one has a complete degeneracy
of all localized magnon states
at the saturation field $h=h_1=\epsilon_1$,
i.e. the energy is $2Ns^2\Delta J-Ns\epsilon_1$
at $h=h_1$
for all $sN-n_{\max}\le S^z\le sN$.
Consequently,
the zero-temperature magnetization $S^z$ jumps
between the saturation value $sN$
and the value
$sN-\frac{1}{6}N$ 
($sN-\frac{1}{9}N$) 
for the square-kagom\'{e} 
(kagom\'{e})
lattice.

Finally,
we mention that the localization of magnon in a finite area 
has some relation to the flat one-magnon dispersion.
Some rigorous results for the flat-band electronic systems 
have been reported in Refs. \cite{13,14,15}.

\section{Spin-Peierls instability}
\label{s2}

To check the lattice stability of the considered systems
with respect to a spin-Peierls mechanism
we assume
a small lattice deformation which preserves the symmetry of the cell
which hosts the localized magnon
(in this case
the independent localized magnon states
remain the exact eigenstates)
and analyze the change in the total energy.
The corresponding deformations
are shown in Fig. \ref{fig01}.
For the square-kagom\'{e} lattice  
(Fig. \ref{fig01}, top)
the deformations  
lead to the following changes in the exchange interactions:
$J\to (1+\sqrt{2}\delta)J$
(along the edges of the square) 
and 
$J\to (1-\frac{1}{4}\sqrt{2}(\sqrt{3}-1)\delta)J$
(along the two edges of the triangles attached to the square),
where the quantity
$\delta$
is proportional to the displacement of the atoms 
and the change in the exchange integrals due lattice distortions 
is taken into account in first order in $\delta$.
For the kagom\'{e} lattice
(Fig. \ref{fig01}, bottom)
one has
$J\to (1+\delta)J$
(along the edges of the hexagon)
and
$J\to (1-\frac{1}{2}\delta)J$
(along the two edges of the triangles attached to the hexagon).
Note, that the lattice distortions shown in Fig. \ref{fig01}
correspond to $\delta>0$.
The magnetic energies (\ref{04}) and (\ref{05})
are changed by distortions
by
$\left(-2\sqrt{2}s+2\sqrt{2}s(2s-1)\Delta
-\frac{\sqrt{3}-1}{\sqrt{2}}s(4s-1)\Delta\right)\delta J$
and
$-s(2+\Delta)\delta J$,
respectively.
Note 
that the linear with respect to $\delta$ change of the magnetic ground-state energies 
has a simple reason:
the magnetic ground-state energy for $sN-n_{\max} \le S^z<sN$ 
contains a sum of contributions coming from local lattice regions 
each of which varies linearly with $\delta$.
The elastic energy increases 
in harmonic approximation by
$2\left(6-\sqrt{3}\right)\alpha\delta^2$ (square-kagom\'{e})
and
by $9\alpha\delta^2$ (kagom\'{e}).
The parameter
$\alpha$
is proportional to the elastic constant of the lattice.
The changes of total energies due to distortions read
\begin{eqnarray}
\label{09}
\Delta{\rm{E}}_1
=
\left(-2\sqrt{2}s+2\sqrt{2}s(2s-1)\Delta
-\frac{\sqrt{3}-1}{\sqrt{2}}s(4s-1)\Delta\right)\delta J
+2\left(6-\sqrt{3}\right)\alpha\delta^2
\end{eqnarray}
and
\begin{eqnarray}
\label{10}
\Delta{\rm{E}}_1
=
-s(2+\Delta)\delta J
+9\alpha\delta^2,
\end{eqnarray}
where for $n$ independent localized magnons
trapped by distorted cells 
these results have to be multiplied by $n$.
Here comes the first important conclusion. 
For the kagom\'{e} lattice 
$\Delta{\rm{E}}_1$ (\ref{10}) decreases at small $\delta>0$ 
for any $\Delta\ge 0$ and any $s$, 
thus
implying a spin-Peierls instability
with the distorted trapping cells 
like in Fig. \ref{fig01}, bottom.
In contrast,
for the square-kagom\'{e} lattice 
with $s>\frac{1}{2}$
the sign of $\delta$ 
which provides a decrease of $\Delta{\rm{E}}_1$ (\ref{09})
for the small lattice distortions 
depends on the anisotropy parameter $\Delta$:
$\delta>0$ if $\Delta<\Delta^\star(s)$
(with  
$\Delta^{\star}(1)\approx 2.21748$,
$\Delta^{\star}(\frac{3}{2})\approx 0.92171$ etc; 
$\Delta^{\star}(s)\sim\frac{1}{s}$ as $s\to\infty$)
but $\delta<0$ if $\Delta>\Delta^\star(s)$.
This can be understood 
as a result of interplay 
between the contributions of transverse and $zz$ correlations along different bonds
for the square-kagom\'{e} lattice geometry.

Let us discuss the scenario of spin-Peierls instability
for the square-kagom\'{e} lattice with $\Delta<\Delta^\star(s)$ 
and 
for the kagom\'{e} lattice.
Consider a magnetic field above the saturation field $h_1$.
For the corresponding
fully polarized ferromagnetic state a lattice distortion is not favorable.
Decreasing $h$ till $h_1$
the homogeneous ferromagnetic state
transforms into the ``distorted magnon crystal''.
Further,
numerical analysis suggests 
that both spin-1 systems exhibit a magnetization plateau
between $h_1$ and $h_2 < h_1$
at $S^z=sN-n_{\max}$.
Calculating the plateau width $\Delta h = h_1-h_2$
for spin-1 isotropic (i.e. $\Delta=1$) finite systems
of 
$N=24,30$
(square-kagom\'{e})
and
$N=27,36,45$
(kagom\'{e})
for the undistorted lattice
($h_2$ is obtained by
$h_2
=E_{\min}(S^z=sN-n_{\max})
-E_{\min}(S^z=sN-n_{\max}-1)$)
and using a $\frac{1}{N}$ finite-size extrapolation
we find some indications
for a finite $\Delta h \approx 0.5 J$ ($\Delta h \approx 0.1 J$)
for the square-kagom\'{e} (kagom\'{e}) lattice
in the thermodynamic limit.
We also mention here
that on the basis of general arguments \cite{16,17} 
one may expect the magnon crystal state to have gapped excitations 
that is related to the magnetization plateau at $S^z=sN-n_{\max}$.

Now the question arises
whether the lattice distortion under consideration
is stable below this plateau,
i.e., for
$S^z < sN-n_{\max}$.
We discuss the question again
for spin-1 finite systems
of size
$N=24$  (square-kagom\'{e})
and
$N=36$ (kagom\'{e})
with $n_{\max}$ distorted squares/hexagons.
We calculate the magnetic energy
for zero and small distortion parameter $\delta$
for different values of $S^z$.
Adopting for the magnetic energy
the ansatz
\begin{eqnarray}
\label{11}
E_{\min}(S^z,\delta)=E_{\min}(S^z,0)+ A\delta^p
\end{eqnarray}
and taking $\delta$ of the order of $10^{-4}$
we estimate the exponent $p$ from the numerical results.
Evidently,
the lattice may become unstable 
if the magnetic energy (\ref{11}) decreases 
with the exponent $p<2$ 
whereas
$p\ge 2$ indicates lattice stability.
Interestingly, 
in the sector of $S^z$ just below the magnon crystal,
$S^z=sN-n_{\max}-1$
(i.e. as $h$ becomes smaller than $h_2$), the 
spin-1 square-kagom\'{e} and kagom\'{e} lattices show different behavior.
For the  kagom\'{e} lattices with $N=36$ we do not find  
a lattice instability. Note that this is in agreement with the situation
for $s=\frac{1}{2}$ \cite{06}. 
Contrary to that,
for the  square-kagom\'{e} lattice with $N=24$ 
we find a lattice instability for $\delta>0$  
also in the sector $S^z=sN-n_{\max}-1=19$ 
in case of small enough anisotropy   
$\Delta<\tilde{\Delta}(1) \approx 1.6$. 
Again if $\Delta>\tilde{\Delta}(1)$
the favorable lattice distortion is characterized by $\delta<0$.
Since we know from Ref. \cite{06}
that for the spin $s=\frac{1}{2}$ square-kagom\'{e} lattice
the spin-Peierls instability can be observed 
even for lower $S^z < sN-n_{\max}-1$,  
we check this also
for the spin $s=1$ square-kagom\'{e} lattice with $N=24$. 
As for $S^z=19$ we find a lattice instability for $\delta>0$ 
only for
small enough anisotropy   
$\Delta \lesssim 1.0$  for $S^z=18$
and $\Delta \lesssim 0.51$ for $S^z=17$, 
i.e. anisotropy parameter $\Delta$
below which this instability occurs 
becomes smaller when $S^z$ 
(i.e. the magnetic field $h$) is diminished.
Thus,
we arrive at the second important conclusion.
For
the spin $s=1$ kagom\'{e} lattice with any $\Delta \ge 0$ 
the spin-Peierls instability 
is favorable only for $sN-n_{\max} \le S^z < sN$
and the distortion disappears for $h<h_2$.
In contrast, 
the distorted spin-1 square-kagom\'{e} lattice 
with the sufficiently small anisotropy parameter $0\le \Delta<\tilde{\Delta}(1)$
remains stable for smaller $h<h_2$.
The spin-1 square-kagom\'{e} lattice 
with $\tilde{\Delta}(1)\le\Delta<\Delta^\star(1)$
exhibits more intricate behavior:
the parameter $\delta$ 
characterizing the lattice distortion at the saturation
changes its sign for $h<h_2$.

As in the spin-$\frac{1}{2}$ case \cite{06}
the saturation field in the distorted lattice is shifted to higher values
that provides a hysteresis phenomenon
in the vicinity of saturation field.

\section{Concluding remarks}
\label{s3}

There is an increasing number of synthesized quantum kagom\'{e} magnets
often with large values of spin
(unfortunately, 
however,
the available at present materials 
are not perfect kagom\'{e} Heisenberg antiferromagnets)
and we may expect that further such materials will be synthesized. 
With our analysis we are pointing out 
that the efforts in this direction are worthwhile 
also because of a new effect:
the spin-Peierls instability in a strong magnetic field.
From the point of view of possible experiments,
it is important to bear in mind the following remarks.
First,
large values of $s$ 
leads to decrease of a relative plateau width
$\frac{\Delta h}{h_1}$.
For instance, 
we have 
$\frac{\Delta h}{h_1}\approx 0.11$
for $s=\frac{1}{2}$ \cite{06}
and 
$\frac{\Delta h}{h_1}\approx 0.08$ 
for $s=1$
(undistorted square-kagom\'{e}, $\Delta=1$);
$\frac{\Delta h}{h_1}\approx 0.023$
for $s=\frac{1}{2}$ \cite{06}
and 
$\frac{\Delta h}{h_1}\approx 0.017$ 
for $s=1$
(undistorted kagom\'{e}, $\Delta=1$).
Thus,
we need materials with not too large $s$
for which the plateau width is not extremely small.
Second, 
we need materials with comparably small exchange constant $J$ 
to have experimentally accessible saturation field $h_1$.
Third,
the presented consideration refers to ideal kagom\'{e} geometry.
We have not discussed the effects of deviation 
from the perfect kagom\'{e} geometry 
on the predicted spin-Peierls instability in high magnetic fields.
Nevertheless 
we may expect 
that the effects of localized magnon states 
will survive 
(see, for example, Ref. \cite{08}).
Fourth,
the reported analysis is performed 
within the frames of the adiabatic treatment of the spin-Peierls instability.
The problem becomes much more difficult 
when the nonadiabatic effects are taken into account.
 
To summarize,
we have examined
a spin-Peierls instability in strong magnetic fields
for two frustrated spin-$s$ anisotropic Heisenberg antiferromagnets
hosting independent localized magnons
demonstrating cumulative effects 
of exchange interaction anisotropy, spin value and lattice geometry.

The present study was supported by the DFG
(Project No. 436 UKR 17/17/03).
O.~D. and J.~R. acknowledge the kind hospitality 
of the MPIPKS, Dresden
in the spring of 2004.
O.~D. is grateful to the ICTP, Trieste
for the kind hospitality 
during the Workshop on Novel States and Phase Transitions in Highly Correlated Matter
(12 July - 23 July, 2004).
The paper was presented 
at the Joint European Magnetic Symposia JEMS`04
(September 5-10, 2004, Dresden, Germany).
O.~D. thanks the Organizers of the JEMS`04  
for the grant for participation in the conference.

\end{document}